\documentclass[12pt]{article}
\usepackage{amsfonts,amsmath,amssymb}
\usepackage{graphicx}

\newfont{\twelvemsb}{msbm10 scaled\magstep1}
\newfont{\eightmsb}{msbm8}
\newfam\msbfam
\textfont\msbfam=\twelvemsb \scriptfont\msbfam=\eightmsb
\catcode`\@=11
\def\Bbb{\ifmmode\let\next\Bbb@\else
\def\next{\errmessage{Use \string\Bbb\space only in math mode}}\fi\next}
\def\Bbb@#1{{\fam\msbfam{{#1}}}}
\begin{document}
\sloppy
\renewcommand{\thefootnote}{\fnsymbol{footnote}}
\newpage
\setcounter{page}{1} \vspace{0.7cm}
\begin{flushright}
June 2007
\end{flushright}
\vspace*{1cm}
\begin{center}
{\bf On the commuting charges for the highest dimension SU(2) operators in planar ${\cal N}=4$ SYM}\\
\vspace{0.5cm} {\large Davide\ Fioravanti $^a$ and Marco\ Rossi}\\
\vspace{.7cm} $^a${\em INFN-Bologna and Department of Physics, University of Bologna,\\
Via Irnerio 46, Bologna, Italy} \\
E-mail: davide.fioravanti@bo.infn.it, rossi@lapp.in2p3.fr
\end{center}
\renewcommand{\thefootnote}{\arabic{footnote}}
\setcounter{footnote}{0} \vspace{3.5cm}
\begin{abstract}
{\noindent We consider the highest anomalous dimension operator in
the $SU(2)$ sector of planar ${\cal N}=4$ SYM at all-loop, though
neglecting wrapping contributions. In any case, the latter enter the
loop expansion only after a precise length-depending order. In the
thermodynamic limit we write both a linear integral equation for the
Bethe root density and a linear system obeyed by the commuting
charges. Consequently, we determine the leading strong coupling
contribution to the density and from this an approximation to the
leading and sub-leading terms of any charge $Q_r$: it scales as
$\lambda ^{1/4-r/2}$, which generalises the Gubser-Klebanov-Polyakov
energy law. In the end, we briefly extend these considerations to
finite lengths and 'excited' operators by using the idea of a
non-linear integral equation.}
\end{abstract}
\vspace{1.5cm}

\newpage

\section {Introduction}

According to the AdS/CFT correspondence \cite {MWGKP}, string theory
on the curved space-time $\text{AdS}_5\times\text{S}^5$ is
equivalent to a conformal quantum field theory on its boundary. In
particular, type IIB superstring theory should be dual to ${\cal
N}=4$ Super Yang-Mills theory (SYM). The AdS/CFT correspondence is a
general dictionary which would in particular equate energies of
string states to anomalous dimensions of local gauge invariant
operators of the quantum field theory.

One of the most important recent development in this context is the
discovery of integrability in both planar field theory and string
theory. In few words, integrable models appear as Bethe equations to
be satisfied by 'rapidities' which parametrise on the one side
composite operators (and their anomalous dimensions in SYM) and on
the other side the corresponding dual objects in string theory, i.e.
states (and their energies, respectively).

In this context, the basic initial result was the identification
\cite{MZ} of the one-loop dilatation operator of scalar
gauge-invariant fields (of bare dimension $L$) with an $SO(6)$
integrable hamiltonian with ($L$ sites). Thanks to this discovery,
one could start to use the powerful technique of the Bethe Ansatz in
order to compute anomalous dimensions of long operators, giving
incredible boost towards a proof of the AdS/CFT correspondence. Soon
afterwards, integrability at higher loops \cite{BKS,SS} was
discovered and an all-loops Bethe Ansatz proposed \cite{BS}. The
restriction to the $SU(2)$ sector of these Bethe equations gives the
previously hinted BDS model \cite{BDS}, which, remarkably, was shown
to be, in a precise sense, the asymptotic limit of the
one-dimensional Hubbard model \cite{RSS, FFGR2}.

In a parallel way, integrability in superstring theory was
discovered first at classical level \cite{BPR} and work at
(semi)classical level allowed to interpret the complex curve
subtending the dynamics in terms of density integral equations,
remembering those in Bethe Ansatz theory \cite {KMMZ}. Moving from
here, the first steps towards extending integrability at quantum
level showed the appearance of long-range Bethe Ansatz equations
with, however, an additional dressing phase \cite{AFS}. More
recently, this phase has been understood as a necessary factor to
guarantee the (non-relativistic) crossing symmetry of the S-matrix
\cite{JAN}. Eventually, a phase factor -- determined by string loop
expansion at first orders \cite{HL} and crossing symmetry -- was
proposed in \cite{BHL,BES}. In \cite {BHL} the complete asymptotic
strong coupling (i.e. string loop) expansion of the dressing factor
was given, whereas in \cite {BES} an expression valid for all values
of the coupling constant was fixed.

In this paper we want to address some of the issues related to the
dressing factor, in a sector, $SU(2)$, different from that, $SL(2)$,
in which it was initially proposed \cite{BES}. First of all, we will
give a proof that the factor given as a meromorphic function of the
coupling constant in \cite{BES} has the asymptotic expansion
proposed by \cite {BHL} \footnote{Indeed, this was not completely
proved in \cite{BES}: we thank M. Staudacher to have encouraged us
on this matter.}. Afterwards, we will write the thermodynamic (i.e.
in the $L\rightarrow \infty$ limit) linear integral equation
satisfied by the density of roots describing the highest (or,
better, 'anti-ferromagnetic') anomalous dimension state and then the
system of linear equations obeyed by (the eigenvalues of) the
commuting charges: we also point out some differences and
difficulties introduced by the presence of the dressing factor.
Then, we find the leading solution to the density equation in the
strong coupling limit and derive, upon integrating this solution, an
approximation to the leading term of any charge. As useful exercise,
we show that in a particular case, corresponding to doubling the
'physical' dressing factor, the aforementioned equations (for both
the density and the charges) allow us for simple solutions in the
strong coupling limit. Therefore, we are able to give the {\it
exact} leading and the sub-leading terms of all the conserved
quantities and we can understand why these contributions are {\it
exact} in this case whereas an approximation in the 'physical' case.
Eventually, we deduce an extension for finite length, $L$, of the
density equation and of the charges equations by using the idea of
the non-linear integral equation \cite{FMQR} along the lines of
\cite{FFGR1}: now the loop expansion can be trusted up to order
$\lambda^{L-1}$. For simplicity's sake we will limit our attention
to the anti-ferromagnetic state and {\it hole} type excitations on
it, even because in other sectors (cf. the treatment of \cite{BES})
these are the only possible states.

\section {Bethe equations in the $SU(2)$ sector}

It is nowadays clear that the {\it asymptotic} Bethe Ansatz type
equations describing planar ${\cal N}=4$ SYM should contain -- with
respect to the first proposals -- a universal dressing phase \cite
{AFS, BHL, BES}. In the $SU(2)$ scalar sector, the BDS equations
\cite {BDS} ought to be modified into
\begin{equation}
\label{BAloops} \left[
\frac{X(u_k+\frac{i}2)}{X(u_k-\frac{i}2)}\right ]^L=
\mathop{\prod^M_{j=1}}_{j\neq k} \frac{u_k-u_j+i}{u_k-u_j-i} \,
{\mbox {exp}}[2i\theta (u_k,u_j)] \, ,
\end{equation}
with the usual notation
\begin{equation}
X(x)=\frac{x}2 \left( 1+\sqrt{1-\frac{\lambda}{4\pi^2 x^2}} \right)
\, . \label{ics}
\end{equation}
On the other hand, the (renormalised) dimension corresponding to the
operator/solution $\{ u_k \}_{k=1,...,M}$ of (\ref {BAloops}) is
formally unchanged, i.e.
\begin{equation}
\Delta = L + \sum _{k=1}^M \left [\left (\frac {1}{2}-iu_k \right)
{\sqrt {1+\frac {4g^2 }{\left (\frac {1}{2}-iu_k \right)^2}}}+ \left
(\frac {1}{2}+iu_k \right) {\sqrt {1+\frac {4g^2 }{\left (\frac
{1}{2}+iu_k \right)^2}}} -1 \right ]  \, , \label {Delta}
\end{equation}
where we have introduced $\lambda=Ng_{YM}^2=16 \pi ^2 g^2$, the 't
Hooft coupling of planar theory ($N\rightarrow \infty,
g_{YM}\rightarrow 0$). The term  {\it asymptotic} means exactly that
this Ansatz is believed to give the exact loop expansion to the
anomalous dimension up to {\it wrapping corrections}, starting at
order $\lambda^L$. The complete phase factor has been conjectured to
be the $\kappa=1$ of the following double series \cite{BES}
\begin{equation}
\theta (u_k,u_j)= \kappa \sum _{r=2}^{\infty} \sum _{\nu
=0}^{\infty} \beta _{r,r+1+2\nu}(g)
[q_r(u_k)q_{r+1+2\nu}(u_j)-q_r(u_j)q_{r+1+2\nu}(u_k)] \, .
\label{theta-beta}
\end{equation}
However, we prefer to keep $\kappa $ unfixed, because in Section 4
we will study the case $\kappa =2$, which mathematically reveals
surprising simplifications. In order to fix the notations in (\ref
{theta-beta}), we remind that $q_r(u)$ is the magnon, $u$, $r$-th
charge \footnote{The following expression may be interpreted as
valid even for the momentum, provided in the sense of the limit
$r\rightarrow 1$.}
\begin{eqnarray}
q_r(u)&=& \left (\frac {8\pi ^2}{\lambda} \right )^{r-1}\frac
{1}{i^{r-2}(r-1)} \left \{ \left [\left (\frac {1}{2}-iu \right)
{\sqrt {1+\frac {4g^2 }{\left (\frac
{1}{2}-iu \right)^2}}}-\left ( \frac {1}{2}-iu\right) \right ]^{r-1} + \right. \nonumber \\
&+& \left. (-1)^r \left [\left (\frac {1}{2}+iu \right) {\sqrt
{1+\frac {4g^2 }{\left (\frac {1}{2}+iu \right)^2}}}-\left ( \frac
{1}{2}+iu\right) \right ]^{r-1} \right \}  \, , \label {qr}
\end{eqnarray}
and the function $\beta _{r,r+1+2\nu}(g)$ are meromorphic functions
of $g$, introduced and studied in \cite {BES}. In this paper their
weak coupling expansion was proposed as
\begin{equation}
\beta _{r,r+1+2\nu}(g)=\sum _{\mu=\nu}^{\infty}
{g}^{2r+2\nu+2\mu}\beta _{r,r+1+2\nu}^{(r+\nu+\mu)} \, ,
\label{betaexp}
\end{equation}
the coefficients $\beta _{r,r+1+2\nu}^{(r+\nu+\mu)}$ being
\begin{equation}
\beta _{r,r+1+2\nu}^{(r+\nu+\mu)}=2(-1)^{r+\mu+1}\frac
{(r-1)(r+2\nu)}{2\mu+1} \left ( \begin{array}  {c} 2\mu +1 \\
\mu-r-\nu+1 \end{array} \right) \left ( \begin{array} {c} 2\mu +1
\\\mu - \nu \end{array} \right ) \zeta (2\mu+1) \, .
\end{equation}
These weak-coupling Taylor series around $g=0$ have finite radius of
convergence and define unambiguously these functions of $g$. In
particular, they can be written in terms of integrals as follows:
\begin{equation}
\beta_{r,r+1+2\nu}(g)=2 (r-1)(r+2\nu) (-1)^{\nu} {g}^{2r+2\nu-1}\int
_0^{\infty} dt \frac {J_{r-1}(2gt)J_{r+2\nu}(2gt)}{t(e^t-1)} \, .
\label{betars}
\end{equation}
Indeed, by expanding the product of Bessel functions in the
integrand of (\ref {betars}) in Taylor series around the origin
\cite{GR}, one can easily check the expansion (\ref{betaexp}).

\medskip

Now we may give a proof that the functions defined by (\ref
{betars}) enjoy at strong coupling ($g\rightarrow +\infty$) the
asymptotic expansion proposed in \cite {BHL}. We reparametrise the
$\beta$s as
\begin{eqnarray}
\beta _{r,s}(g)&=&g^{r+s-2}c_{r,s}(g) \, , \nonumber \\
c_{r,s}(g)&=& 2 \cos \left [ \frac {\pi (s-r-1)}{2} \right ]
(r-1)(s-1) \int _0^{\infty} dt \frac
{J_{r-1}(2gt)J_{s-1}(2gt)}{t(e^t-1)} \, , \label{beta-c}
\end{eqnarray}
in order to make contact with the notations of \cite {BHL}. From
their definition it is always $r\geq 2$, $s\geq r+1$ and their
difference $s-r$ equals a positive odd integer. We now perform the
change of variables $t\rightarrow t/2g$ and expand the function
$1/(e^{t/2g}-1)$ in powers of $t/2g$. Then we formally exchange this
power series with the integral, obtaining
\begin{equation}
\int _0^{\infty} dt \frac {J_{r-1}(2gt)J_{s-1}(2gt)}{t(e^t-1)} =
\sum _{n=0}^{\infty} \frac {1}{(2g)^{n-1}}\frac {B_n}{n!} \int
_{0}^{\infty} dt  J_{r-1}(t)J_{s-1}(t)  t^{n-2} \, , \label {bern}
\end{equation}
with $B_n$ the Bernoulli numbers. The series in (\ref {bern}) is for
now formal, since the integral is defined only for $n \leq 2$.
However, the idea is to extend the result obtained for general $n$,
$0\leq n \leq 2$, to arbitrary values of $n$. This will eventually
define the asymptotic expansion of the coefficients $c_{r,s}(g)$.
Indeed, we may specialise formula 6.574.2 of \cite {GR} to the form
\begin{equation}
\int _0^{\infty} dt J_{r-1}(t)J_{s-1}(t) t^{n-2}=\frac
{2^{n-2}\Gamma (-n+2) \Gamma \left ( \frac {r+s-3+n}{2}\right )
}{\Gamma \left ( \frac {s-r+3-n}{2}\right )\Gamma \left ( \frac
{r+s+1-n}{2}\right )\Gamma \left ( \frac {r-s+3-n}{2}\right )} \, .
\label {Jint}
\end{equation}
Strictly speaking this formula is valid for $n=0,1$. When $n=0$ the
last Gamma function in the denominator is divergent unless $s=r+1$,
and then
\begin{equation}
\int _0^{\infty} dt J_{r-1}(t)J_{s-1}(t) t^{-2}=\delta _{s,r+1}\frac
{1}{4r(r-1)} \, .
\end{equation}
For $n=1$ the expression may be simplified into
\begin{equation}
\int _0^{\infty} dt J_{r-1}(t)J_{s-1}(t) t^{-1}=\frac {2 \sin \left
(\pi \frac {s-r}{2} \right )}{\pi} \frac {1}{(s+r-2)(s-r)} \, .
\end{equation}
For what concerns the values $n\geq 2$, we remark that the right
hand side of (\ref {Jint}) can be well defined not only for $n=2$,
but also for all even $n$. In order to prove this, it is useful to
introduce the integer $k\geq 0$, such that $s-r=2k+1$, and the
regularisator $\delta \rightarrow 2$
\begin{equation}
\int _0^{\infty} dt J_{r-1}(t)J_{s-1}(t) t^{n-\delta}=\frac
{2^{n-\delta}\Gamma (-n+\delta) \Gamma \left ( k+r+\frac
{n-\delta}{2}\right ) }{\Gamma \left ( k+1-\frac {n-\delta}{2}\right
)\Gamma \left (  k+r-\frac {n-\delta}{2}\right )\Gamma \left (
-k-\frac {n-\delta}{2}\right )}.
\end{equation}
In this limit, the first Gamma function in the numerator as well as
the last in the denominator diverge, but their ratio stays finite:
\begin{equation}
\lim _{\delta \rightarrow 2}\frac {\Gamma (\delta -n)}{\Gamma \left
(-k +\frac {\delta -n}{2} \right )}= \frac {1}{2}(-1)^\frac
{n-s+r-1}{2} \frac {\Gamma \left ( \frac {s-r-1+n}{2}\right)
}{\Gamma (n-1)} \, .
\end{equation}
Therefore, we may write
\begin{equation}
\int _0^{\infty} dt J_{r-1}(t)J_{s-1}(t) t^{n-2}=(-1)^\frac
{n-s+r-1}{2} \frac {2^{n-3}\Gamma \left (\frac {s-r-1+n}{2}\right )
\Gamma \left (\frac {s+r-3+n}{2}\right )}{\Gamma (n-1) \Gamma \left
(\frac {s-r+3-n}{2}\right ) \Gamma \left (\frac {s+r+1-n}{2}\right )
} \, , \label{JJint}
\end{equation}
for all $n$ even, $n\geq 2$. Because of the divergence of the second
Gamma function in the denominator, the above result is different
from zero only if
\begin{equation}
n\leqslant 2k+2=s-r+1 \, .
\end{equation}
We now plug (\ref {JJint}) into (\ref {bern}) and obtain the
coefficients $c_{r,s}^{(n)}$ of the asymptotic expansion
\begin{equation}
c_{r,s}(g)=\sum _{n=0}^{\infty} c_{r,s}^{(n)}g^{1-n} \, , \label
{cstrong}
\end{equation}
in the form
\begin{equation}
c_{r,s}^{(0)}=\delta _{r+1,s} \, , \quad c_{r,s}^{(1)}=-\frac
{2}{\pi} \frac {(r-1)(s-1)}{(s+r-2)(s-r)} \, . \label {c01}
\end{equation}
and for $n\geq 2$
\begin{equation}
c_{r,s}^{(n)}=\frac {1}{(-2\pi)^n \Gamma (n-1)}\zeta (n)
(r-1)(s-1)\frac {\Gamma \left (\frac {s+r+n-3}{2}\right ) \Gamma
\left (\frac {s-r+n-1}{2}\right )}{ \Gamma \left (\frac
{s+r-n+1}{2}\right ) \Gamma \left (\frac {s-r-n+3}{2}\right )} \, .
\label{crsn}
\end{equation}
In fact, this expression for any integer $n\geq 2$ may be obtained
upon expressing the Bernoulli number $B_n$ via the Riemann zeta
function $\zeta (n)$, even for odd $n$, as \footnote{It may be
proved that this amounts to a further regularisation of the
coefficients of the series (\ref{bern}).}
\begin{equation}
B_n=(-1)^{\frac {n}{2}-1}\frac {\zeta (n) n!}{2^{n-1}\pi ^n} \, .
\end{equation}

To summarise, we have obtained the strong coupling expansion
(\ref{cstrong}-\ref {crsn}) for the functions $c_{r,s}(g)$ entering
the dressing factor. Notice that in formul{\ae} (\ref {c01},\ref
{crsn}) $r\geq 2$ and $s-r$ equals a positive odd integer and
(\ref{crsn}) implies when $n$ is even $c_{r,s}^{(n)}=0$ for
$n>s-r+1$.

\section {Thermodynamic limit of the highest energy state}
\setcounter{equation}{0}

In the infinite length limit $L\rightarrow \infty$, there is no
possibility for wrapping and consequently the equations
(\ref{BAloops}) should give an exact description of the $SU(2)$
scalar sector. In this limit, the Bethe roots become closer and
closer in such a way to form a continuous distribution. The latter
may be described by a density function $\rho(u)$ and the Bethe
equations turn into a linear integral equation.

In order to find this equation, we rewrite equations (\ref
{BAloops}) in the logarithmic form,
\begin{equation}
\label{logBAloops} iL\ln \frac{X\left (u_k+\frac{i}{2}\right )}
{X\left (u_k-\frac{i}{2}\right )}= i\mathop{\sum ^M_{j=1}}_{j\neq k}
\ln \frac{u_k-u_j+i}{u_k-u_j-i} - 2 \mathop {\sum ^M_{j=1}}_{j\neq
k} \theta (u_k,u_j) +2\pi K_k\, ,
\end{equation}
where $K_k$ are integers which depend on the state we are
considering. Now, for simplicity's sake we specialise our analysis
to the state with uniquely $M=L/2$ real roots (no complex roots).
This corresponds to the anti-ferromagnetic or highest anomalous
dimension configuration and there is not even room for holes and
then $K_k=k$. When $L\rightarrow \infty$ all the sums over Bethe
roots $u=u_k$ may be replaced by integrals with Stieltjes measure
\begin{equation}
du\rho (u)=du \frac {1}{L}\frac {dk}{du} \, , \label{rhodef}
\end{equation}
and hence, upon derivating with respect to $u$, each term
(\ref{logBAloops}) may be expressed in terms of $\rho$ itself
\begin{eqnarray}
&& i \frac {X^{\prime}\left (u +\frac {i}{2} \right)}{X\left (u
+\frac {i}{2} \right)} -i \frac {X^{\prime}\left (u -\frac {i}{2}
\right)}{X\left (u -\frac {i}{2} \right)}=i \int _{-\infty}^{\infty}
dv \rho (v) \left [ \frac
{1}{u-v+i}- \frac {1}{u-v-i} \right ] - \nonumber \\
&-& 2  \int _{-\infty}^{\infty} dv \rho (v) \frac {d}{du} \theta
(u,v) +2\pi \rho (u) \, . \nonumber
\end{eqnarray}
Inserting the form (\ref {theta-beta}) for the dressing phase, we
have
\begin{eqnarray}
&& \frac {i}{2}\left [ \frac {1}{\sqrt {\left (u+\frac
{i}{2}\right)^2 -4g^2 }}-
 \frac {1}{\sqrt {\left (u-\frac {i}{2}\right)^2 -4g^2 }} \right ] = \pi
\rho (u) +
\int _{-\infty}^{\infty} dv \frac {\rho (v)}{(u-v)^2 +1} - \nonumber \\
&-& \kappa \sum _{r=2}^{\infty} \sum _{\nu =0}^{\infty} \beta
_{r,r+1+2\nu}(g) q_r^{\prime}(u) \int _{-\infty}^{\infty} dv \rho
(v)
q_{r+1+2\nu}(v) + \nonumber  \\
&+& \kappa \sum _{r=2}^{\infty} \sum _{\nu =0}^{\infty} \beta
_{r,r+1+2\nu}(g) q_{r+1+2\nu}^{\prime}(u) \int _{-\infty}^{\infty}
dv \rho (v) q_{r}(v) \, . \nonumber
\end{eqnarray}
In order to have shorter expressions, we introduce the (total)
charges, normalised by a factor $1/L$,
\begin{equation}
Q_r=\int _{-\infty}^{\infty} dv \rho (v) q_{r}(v) \, . \label{Qr}
\end{equation}
As usual, we pass to the Fourier transform, using the simple
expression \cite{GR}
\begin{equation}
\hat q_r(k)=2^{r-1} \frac {(2\pi)^r} { (\sqrt {\lambda})^{r-1}}
\frac {1}{i^{r-2}}\frac {J_{r-1} \left (\frac {\sqrt
{\lambda}}{2\pi} k \right ) }{k e^{\frac {|k|}{2}}} \, , \quad r
\geq 1 \, . \label {qrk}
\end{equation}
Then, we obtain for the Fourier transform of the density
\begin{eqnarray}
&& \pi e^{-\frac {|k|}{2}} J_0\left ( \frac {\sqrt {\lambda}}{2\pi}
k\right )=\pi \hat \rho (k) + \pi \hat \rho (k) e^{-|k|} - \nonumber \\
&-&  \kappa \sum _{r=2}^{\infty} \sum _{\nu =0}^{\infty} \beta
_{r,r+1+2\nu}(g) 2^{r-1} \frac {(2\pi)^r} { (\sqrt {\lambda})^{r-1}}
\frac {1}{i^{r-3}} e^{-\frac {|k|}{2}}J_{r-1} \left (\frac {\sqrt
{\lambda}}{2\pi} k \right )
Q_{r+1+2\nu} + \nonumber \\
&+& \kappa \sum _{r=2}^{\infty} \sum _{\nu =0}^{\infty} \beta
_{r,r+1+2\nu}(g) 2^{r+2\nu} \frac {(2\pi)^{r+1+2\nu}} { (\sqrt
{\lambda})^{r+2\nu}} \frac {1}{i^{r-2+2\nu}} e^{-\frac
{|k|}{2}}J_{r+2\nu} \left (\frac {\sqrt {\lambda}}{2\pi} k \right )
Q_{r} \, . \nonumber
\end{eqnarray}
Collecting the terms containing $\hat \rho (k)$ all together, we may
deduce a linear integral equation
\begin{eqnarray}
\hat \rho (k)&=&\frac {J_0\left ( \frac  {\sqrt {\lambda}}{2\pi} k
\right)}{2 \cosh \frac {k}{2}}+ \kappa
 \sum _{r=2}^{\infty} \sum _{\nu =0}^{\infty} \beta _{r,r+1+2\nu}(g)
2^{r-1} \frac {(2\pi)^r} { (\sqrt {\lambda})^{r-1}} \frac
{1}{i^{r-3}} \frac {J_{r-1} \left (\frac {\sqrt {\lambda}}{2\pi} k
\right )}{2 \pi
\cosh \frac {k}{2}} Q_{r+1+2\nu} - \nonumber \\
&-&  \kappa \sum _{r=2}^{\infty} \sum _{\nu =0}^{\infty} \beta
_{r,r+1+2\nu}(g) 2^{r+2\nu} \frac {(2\pi)^{r+1+2\nu}} { (\sqrt
{\lambda})^{r+2\nu}} \frac {1}{i^{r-2+2\nu}} \frac {J_{r+2\nu} \left
(\frac {\sqrt {\lambda}}{2\pi} k \right )}{2 \pi \cosh \frac {k}{2}}
Q_{r}  \label {rhoeq1} \, .
\end{eqnarray}
Of course, this equation may entails an analogous linear equation
for the density of roots $\rho(u)$. But more importantly it tells us
that the charges $Q_r=Q_r(g)$ determine $\hat\rho(k) $ for any $g$;
and on its turn $\hat\rho(k) $ yields the charges (\ref{Qr}) as
\begin{equation}
Q_r=\int _{-\infty}^{\infty} \frac {dk}{2\pi} \hat \rho (k) \hat
q_r(-k)=\int _{-\infty}^{\infty}  dk \hat \rho (k) \frac
{i^{r-2}}{g^{r-1}} \frac {J_{r-1} ( 2 g k)}{k e^{\frac {|k|}{2}}} \,
.   \label {Qrk}
\end{equation}
Thanks to this sort of r\^ole exchange, we think that the form of
the integral equation in terms of the charges is of particular
interest. As an example of this utility, we will see in the
following the derivation of a linear system of algebraic equations
for the charges.

In fact, upon inserting (\ref {Qrk}) in (\ref {rhoeq1}), we may
rewrite the equation for $\hat \rho (k)$ in a more explicit way
\begin{eqnarray}
\hat \rho (k)&=&\frac {J_0( 2 g k)}{2 \cosh \frac {k}{2}}+
 \kappa \sum _{r=2}^{\infty} \sum _{\nu =0}^{\infty} \beta _{r,r+1+2\nu}( g)
(-1)^{1+\nu}{g}^{1-2\nu -2r}  \Bigl [ \frac {J_{r-1}  (2 g k
)}{\cosh
\frac {k}{2}} \cdot \nonumber \\
&\cdot & \int _{-\infty}^{\infty} dp \hat \rho (p) \frac {J_{r+2\nu}
( 2 g p)}{p e^{\frac {|p|}{2}}} + \frac {J_{r+2\nu} (2 g k  )}{\cosh
\frac {k}{2}} \int _{-\infty}^{\infty} dp \hat \rho (p) \frac
{J_{r-1} ( 2 g p)}{p e^{\frac {|p|}{2}}} \Bigr ] \, , \label
{rhoeq2}
\end{eqnarray}
still slightly simplified if in terms of the $c_{r,s}(g)$ (using
(\ref {beta-c})):
\begin{eqnarray}
\hat \rho (k)&=&\frac {J_0( 2 g k)}{2 \cosh \frac {k}{2}}+
 \kappa \sum _{r=2}^{\infty} \sum _{\nu =0}^{\infty} c_{r,r+1+2\nu}(g)
(-1)^{1+\nu} \Bigl [ \frac {J_{r-1}  (2 g k )}{\cosh \frac {k}{2}}
\cdot
\nonumber \\
&\cdot & \int _{-\infty}^{\infty} dp \hat \rho (p) \frac {J_{r+2\nu}
( 2 g p)}{p e^{\frac {|p|}{2}}} + \frac {J_{r+2\nu}  (2 g k )}{\cosh
\frac {k}{2}} \int _{-\infty}^{\infty} dp \hat \rho (p) \frac
{J_{r-1} ( 2 g p)}{p e^{\frac {|p|}{2}}} \Bigr ] \, . \label
{rhoeq3}
\end{eqnarray}
In the end, upon highlighting in this equation, via (\ref{Qrk}), the
explicit dependence on the conserved charges, we may derive
immediately an infinite set of linear equations for them:
\begin{eqnarray}
&&Q_s=\frac {i^{s-2}}{g^{s-1}} \Bigl [ \int _{-\infty}^{\infty} dk
\frac {J_{s-1}
( 2 g k) J_0( 2 g k)}{k (e^{|k|}+1)} + \nonumber \\
&+& 2\kappa  \sum _{r=2}^{\infty} \sum _{\nu =0}^{\infty}
c_{r,r+1+2\nu}(g) (-1)^{1+\nu} \int _{-\infty}^{\infty} dk \frac
{J_{s-1} ( 2 g k) J_{r-1}( 2 g k)}{k (e^{|k|}+1)}  \frac {
g^{r+2\nu}}{
i^{r+2\nu-1}} Q_{r+2\nu+1}  \nonumber \\
&+&  2\kappa  \sum _{r=2}^{\infty} \sum _{\nu =0}^{\infty}
c_{r,r+1+2\nu}(g) (-1)^{1+\nu} \int _{-\infty}^{\infty} dk \frac
{J_{s-1} ( 2 g k) J_{r+2\nu}( 2 g k)}{k (e^{|k|}+1)} \frac {
g^{r-1}}{i^{r-2}} Q_{r} \Bigr ] \, . \label{eqcharg}
\end{eqnarray}
We want to remark that this equations form a system of linear
algebraic equations for (the eigenvalues of) the commuting charges
on the highest energy state. We believe it could furnish important
information on these eigenvalues and could introduce their
disentanglement under particular conditions (for instance in the
strong coupling regime).

We have to remark that an equivalent equation for $\hat \rho(k)$ has
been already published in \cite{RSZ} by making use of their
formalism of {\it magic kernels} \cite{BES}.

\medskip

Both equations (\ref {rhoeq3}) and (\ref {eqcharg}) characterise the
$L\rightarrow \infty$ limit of the highest anomalous dimension state
of the scalar sector of ${\cal N}=4$ SYM, once one supposes this to
be described by the Ansatz (\ref {BAloops}). One has to remark the
important modifications due to the presence of the dressing phase.
Without that phase (i.e. in the framework of the BDS Bethe Ansatz),
the thermodynamic expressions for $\hat \rho (k)$ and $Q_s$ reduce
to
\begin{eqnarray}
\hat \rho (k)&=&\frac {J_0( 2 g k)}{2 \cosh \frac {k}{2}} \, , \nonumber \\
Q_s&=&\frac {i^{s-2}}{g^{s-1}} \int _{-\infty}^{\infty} dk \frac
{J_{s-1} ( 2 g k) J_0( 2 g k)}{k (e^{|k|}+1)} \, . \label{BDSsol}
\end{eqnarray}
By construction, at weak coupling the dressing factor starts at
order $g^6$ and therefore, up to that order, the solutions of (\ref
{rhoeq3}) and (\ref {eqcharg}) are given by (\ref {BDSsol}). On the
other hand, the strong coupling limit is dominated by contributions
coming from the dressing factor and it is of great importance, since
it makes contact with semiclassical results in string theory (cf.
second reference of \cite{MWGKP}).

\medskip

We remark the presence in equations (\ref {eqcharg}) of the
quantities
\begin{equation}
\tilde c_{r,s}(g)=\int _0^{\infty} dt \frac
{J_{r-1}(2gt)J_{s-1}(2gt)}{t(e^t+1)} \, , \label{tildec}
\end{equation}
which look very similar to the functions $c_{r,s}( g)$ appearing in
the dressing factor. Actually, the elementary identity
\begin{equation}
\frac {1}{e^x-1}=\frac {1}{2}\left (\frac {1}{e^{\frac {x}{2}}-1}-
\frac {1}{e^{\frac {x}{2}}+1}\right )
\end{equation}
allows us to write, when $r \geq 2$ and $s-r$ equals a positive odd
integer,
\begin{equation}
2 \tilde c_{r,s}(g)= \frac {c_{r,s}(g) -2 c_{r,s}\left (\frac
{g}{2}\right)}{\cos \left [ \frac {\pi (s-r-1)}{2} \right ]
(r-1)(s-1) } \, . \label{cctilda}
\end{equation}
This entails the strong coupling expansion
\begin{equation}
\tilde c_{r,s} (g)=\sum _{n=0}^{\infty} \tilde c_{r,s}^{(n)}
{g}^{1-n} \,
\end{equation}
of the integrals contained in (\ref {eqcharg}) in terms of the
coefficients (\ref {c01},\ref {crsn}):
\begin{equation}
\tilde c_{r,s}^{(n)}=\frac {(1-2^n)c_{r,s}^{(n)}}{2(r-1)(s-1)\sin
\frac {\pi (s-r)}{2}}\, .
\end{equation}
In particular, we get $\tilde c_{r,s}^{(0)}=0$ and
\begin{eqnarray}
\tilde c_{r,s}^{(1)} &=&\frac {1}{\pi}\frac {1}{(s+r-2)(s-r)}\frac
{1}{\sin \frac {\pi (s-r)}{2}} \, ; \label {tilc1} \\
\tilde c_{r,s}^{(n)}&=&\frac {1}{(-2\pi)^n \Gamma (n-1)}\zeta (n)
\frac {(1-2^n)}{2 \sin \frac {\pi (s-r)}{2}} \frac {\Gamma \left
(\frac {s+r+n-3}{2}\right ) \Gamma \left (\frac {s-r+n-1}{2}\right
)}{ \Gamma \left (\frac {s+r-n+1}{2}\right ) \Gamma \left (\frac
{s-r-n+3}{2}\right )} \, . \label {tilcn}
\end{eqnarray}
We remember that results (\ref {tilc1}, \ref {tilcn}) are obtained
when $r\geq 2$ and $s-r$ is equal to a positive odd integer. They
can be extended to other values of $r,s$ by using the symmetry
property $\tilde c_{r,s}(g)=\tilde c_{s,r}(g)$. Actually, this is
what we will need for applications in this paper.

As a final remark, we observe that, with the before-stated
restrictions on $r$ and $s$, relation (\ref{cctilda}) may be
inverted as
\begin{equation}
c_{r,s}(g)=2 \cos \left [ \frac {\pi (s-r-1)}{2} \right ] (r-1)(s-1)
\sum _{n=0}^{\infty} 2^n \tilde c_{r,s}(2^{-n} g) \, .
\end{equation}

\section{Strong coupling limit}
\setcounter{equation}{0}

It is of great interest to find the solutions of the linear
equations for ${\hat \rho}$ and the charges $Q_r$ in the strong
coupling limit (i.e. $g \rightarrow +\infty$), as these ought to
match the semiclassical string theory expansion (provided no
attention is payed to the order of the two limits
$L\rightarrow\infty$ and $g\rightarrow+\infty$ as shown in
\cite{FFGR2}). Referring to formula (\ref {theta-beta}), we will
first concentrate on the case of generic $\kappa $, where we will be
able to find the leading term for $\rho $ and, in a sense we will
specify afterwards, to provide information on the leading term of
the charges $Q_r$. In a second moment, we will consider the
particular case $\kappa =2$, where, because of considerable
mathematical simplifications, it is possible to find exactly the
leading terms of the density $\rho $ and of the eigenvalues of all
conserved charges $Q_r$.

We have to remark that in the final proposal of \cite{BES} the value
of interest is $\kappa =1$. In this case, numerical computation of
the strong coupling limit of the anomalous dimension was first
performed in \cite {BD} and, subsequently, analytical confirmation
of their finding was given in \cite {KSV}, following, though, a way
different from ours. Nonetheless, we would like here to analyse the
strong coupling expansion of ${\hat \rho}$ and of the charges, and
the difficulties in their comparison.

\subsection{Generic $\kappa$}

Equation (\ref{rhoeq1}) clearly suggests and gives a full meaning to
what was proposed in \cite{AABEK}, i.e. that $2\cosh \frac {k}{2} \
\hat \rho (k)$ may be developed in series of Bessel functions
\begin{equation}
2\cosh \frac {k}{2} \ \hat \rho (k)= \sum _{m=0}^{\infty}
a_{2m}(g)J_{2m}(2gk) \, . \label{rhoexp}
\end{equation}
We insert this expansion in (\ref {rhoeq3}) and re-write (\ref
{rhoeq3}) as a linear system of equations for the coefficients
$a_{2n}(g)$\footnote {We are supposing that it is possible to
exchange the symbol of integral with that of series: in general, we
have no specific guarantee of the validity of this hypothesis. On
the contrary, counterexamples are known; however, it is customary
(cf. for instance \cite{AABEK}) to operate the exchange at this
stage.}:
\begin{eqnarray}
a_0(g)&=&1 \, \nonumber \\
a_{2n}(g)&=& 4 \kappa \Bigl [ \sum _{\nu=0}^{\infty} \sum
_{m=0}^{\infty} (-1)^{1+\nu} c_{2n+1,2n+2\nu+2}(g)\tilde
c_{2m+1,2n+2\nu+2}(g)\, a_{2m}(g) + \label {eqmatr} \\
&+& \sum _{\nu=0}^{n-1} \sum _{m=0}^{\infty} (-1)^{1+\nu}
c_{2n-2\nu,2n+1}(g)\tilde c_{2n-2\nu,2m+1}(g)\, a_{2m}(g) \Bigr ] \,
, \quad n\geq 1 . \nonumber
\end{eqnarray}
We take now the limit $g\rightarrow +\infty$. Supposing that, for $n
\geq 1$,
\begin{equation}
\lim _{g\rightarrow +\infty} a_{2n}(g)=a_{2n}^{(0)} \, ,
\end{equation}
the leading order $O(g)$ of the second of equations (\ref {eqmatr})
is \footnote{Here we are again exchanging the above limit (or the
corresponding asymptotic series) with the series. This may be
troublesome for the next-to-leading correction to $a_{2n}(g)$.}
\begin{equation}
\tilde c_{1,2n+2}^{(1)} + \tilde c_{2n,1}^{(1)}+
\sum_{m=1}^{\infty}[\tilde c_{2m+1,2n+2}^{(1)}+\tilde
c_{2n,2m+1}^{(1)}]\, a_{2m}^{(0)}=0 \, ,
\end{equation}
where $\tilde c_{r,s}^{(1)}$ is given by (\ref {tilc1}). The
solution of this equation (remember that $a_0^{(0)}=1$) is
\begin{equation}
a_{2m}^{(0)}=2(-1)^m \, , \quad m \geq 1 \, .
\end{equation}
Therefore, for the leading term of the density,
\begin{equation}
\hat \rho ^{(0)} (k)= \frac {1}{2\cosh \frac {k}{2}} \sum
_{n=0}^{\infty} a_{2n}^{(0)}J_{2n}(2gk) \, , \label {rhoexplea}
\end{equation}
we obtain through summation formul{\ae} for Bessel functions in
\cite {GR}, the expression
\begin{equation}
\hat \rho ^{(0)} (k)= \frac {\cos ( 2 g k)}{2 \cosh \frac {k}{2}} \,
\,  \Rightarrow \rho ^{(0)} (u) = \frac {1}{4} \left [ \frac
{1}{\cosh \pi (u+ 2 g)}+ \frac {1}{\cosh \pi (u- 2 g)} \right ] \, .
\label {rhozerolimit}
\end{equation}
This form for the leading density is not surprising. Numerical
considerations, presented in \cite {BD} for $\kappa =1$, show that
the momentum of one {\it magnon} in the anti-ferromagnetic state
scales, at strong coupling, as
\begin{equation}
p_k = \alpha _k \lambda ^{-\frac {1}{4}}+.... \, .
\end{equation}
This implies that the rapidity $u_k=\frac {1}{2} {\mbox {cotg}}
\frac {p_k}{2} \sqrt { 1+\frac {\lambda}{\pi ^2}\sin ^2  \frac
{p_k}{2}}$ behaves like
\begin{equation}
u_k =\frac {{\mbox {sgn}}(\alpha _k)}{2\pi}\lambda ^{\frac
{1}{2}}+...\, .
\end{equation}
Therefore, Bethe roots accumulate at the two values $\pm \frac
{\lambda^{\frac {1}{2}}}{2\pi}=\pm 2 g$, exactly as suggested by
(\ref {rhozerolimit}).

Yet, owing to the presence of $g$ in the argument of the Bessel
functions in the series (\ref {rhoexplea}), the leading expression
for $\hat \rho (k)$ (\ref{rhozerolimit}) can be really trusted only
for $gk \simeq 1$ (and $g\gg 1$). In particular, this means that, if
we insert (\ref {rhozerolimit}) in (\ref {Qr}), we only obtain an
approximation for the leading value of (the eigenvalues of) the
charges $Q_r$. For even $r$ (if $r$ is odd, we have $Q_r=0$, since
$\rho (u)=\rho (-u)$) we may indeed conclude
\begin{eqnarray}
&&Q_r\simeq \frac {2^{2-r}}{4i^{r-2}(r-1)g^{2r-2}}\left \{ \int
_{-\infty}^{\infty} \frac {du}{\cosh \pi u} \left [ \left ( \frac
{1}{2}-iu+2 i g \right ){\sqrt {1 +\frac {4g^2} {\left ( \frac
{1}{2}-iu+2i g \right )^2}}} -  \right. \right. \nonumber \\
&-& \left. \left. \frac {1}{2} + iu- 2 i g  \right] ^{r-1} + h.c.
\right \}=  \frac {2^{-\frac {5}{2}}}{g^{r-\frac {1}{2}}}I + O\left
( \frac {1}{g^r} \right )
  \, .  \label {Qrgk}
\end{eqnarray}
We will go on expanding in the following, but for now we just
concentrate on the first term where the integral
\begin{equation}
I=2 \int _{-\infty}^{\infty} \frac {du}{\cosh \pi u} [ {\sqrt
{i+2u}}+{\sqrt {-i+2u}}]
\end{equation}
may be computed by moving the domain of integration on the line
Im$u=-1/2$ i.e. changing variable $y=\frac {\pi}{2}(i+2u)$:
\begin{equation}
I=\frac {4 {\sqrt {2}}}{\pi ^{\frac {3}{2}}}\int _{0}^{\infty} dy
\frac {\sqrt {y}}{\sinh y}=\frac {2}{\pi}(2^{\frac {3}{2}}-1) \zeta
\left ( \frac {3}{2} \right ) =3.04084....\, . \label {Ivalue}
\end{equation}
As expected, our result, which does not depend on $\kappa$, differs
from the numerical computation of \cite{BD}, valid for $\kappa =1$:
in formula (5.7) of their paper they state that
\begin{equation}
\Delta = L \left [ \pi ^{\frac {1}{2}} g^{\frac {1}{2}} + O(g^0)
\right ] \, , \label {BDdelta}
\end{equation}
which would come out from a value $I=2 {\sqrt {2\pi}}$.
Nevertheless, we may reproduce the 'right' value (\ref{BDdelta}) by
shifting the accumulation points in the limiting delta-functions,
i.e.
\begin{equation}
\rho ^{(0)}(u)=\frac {1}{4} \left [ \delta \left (u+2g-\frac
{\pi}{4}+\frac {1}{4\pi} \right)+
 \delta \left (u-2g+\frac {\pi}{4}-\frac {1}{4\pi} \right) \right ] \, .
\end{equation}
These shifts might allow for a simple interpretation as an effect of
sub-leading corrections to $\hat \rho (k)$, though the expression
before is not the leading term (in the sense that has been defined
above).

\subsection{The case $\kappa =2$}

The $\kappa =2$ case reveals surprising mathematical
simplifications, as here the formul{\ae} above for the density and
(the eigenvalues of) the charges ((\ref{rhozerolimit}) and
(\ref{Qrgk}) respectively) turn out to be the {\it exact} leading
terms without exchanging limit processes. For this reason, we
believe this case may be important to understand in a deeper manner
the meaning of the strong coupling expansion\footnote{This exactness
do not exclude, on the contrary may support, the need for changing
the expansion as defined by (\ref{rhoexp}).}.

As discussed in the previous subsection, Bethe roots accumulate at
$\pm 2g$ independently of the value of $\kappa$. Therefore, their
density $\rho (u)$ ought to have the form
\begin{equation}
\rho (u)=\frac {1}{4} \left [ f(u+ 2 g)+ f(-u+2 g) \right ] \, ,
\label {rholimit}
\end{equation}
where $f(u)$ is a function peaked around $u=0$, not depending on $g$
and such that
\begin{equation}
\int _{-\infty}^{\infty} du f(u) =1 \, .
\end{equation}
Now, we show that the guess (\ref{rholimit}), with a proper choice
of $f$, is indeed solution of the linear equation (\ref{rhoeq3})
when $g\rightarrow +\infty $. In fact, this form allow us to compute
the charges $Q_r$ appearing in (\ref{rhoeq2}). If $r$ is odd, we
have $Q_r=0$, since $\rho (u)=\rho (-u)$. Otherwise, for even $r$,
we gain
\begin{eqnarray}
&&Q_r=\frac {2^{2-r}}{4i^{r-2}(r-1)g^{2r-2}}\left \{ \int
_{-\infty}^{\infty} {du}f(u) \left [ \left ( \frac {1}{2}-iu+2 i g
\right ){\sqrt {1 +\frac {4g^2} {\left ( \frac {1}{2}-iu+2 i g
\right )^2}}} -
\right. \right. \nonumber \\
&-& \left. \left. \frac {1}{2} + iu- 2 i g  \right] ^{r-1} + h.c.
\right \}= \left [ \frac {2^{-\frac {5}{2}}}{g^{r-\frac {1}{2}}}I_f-
(r-1)\frac {2^{-2}}{g^r} +O\left (\frac {1}{g^{r+\frac {1}{2}}}
\right ) \right ]  \, ,  \label {Qrg}
\end{eqnarray}
where the leading term is proportional to the f-depending integral
\begin{equation}
I_f=2 \int _{-\infty}^{\infty} {du} f(u) [ {\sqrt {i+2u}}+{\sqrt
{-i+2u}}] \, , \label {I}
\end{equation}
whereas the sub-leading term does not depend on $f$.

\medskip

Now, we shall plug the result (\ref{Qrg}) into the linear equation
(\ref {rhoeq3}) in which we have consistently developed the
coefficients $c_{r,s}(g)$ according to (\ref{crsn}): in this
respect, it is sufficient to insert the leading contribution
$c_{r,s}( g)=\delta _{r+1,s} \, g$:
\begin{eqnarray}
\hat \rho (k)&=&\frac {J_0( 2 g k)}{2 \cosh \frac {k}{2}}- \nonumber \\
&-& 4g \sum _{m=1}^\infty \frac {J_{2m}( 2 g k)}{ \cosh \frac
{k}{2}} \frac {g^{2m+1}}{2(-1)^m}\left [ \frac {2^{-\frac
{5}{2}}}{g^{2m+\frac {3}{2}}}I_f- (2m+1)\frac {2^{-2}}{g^{2m+2}}
\right ] [1+O(g^{-1})] -
\nonumber \\
&-& 4 g\sum _{m=1}^\infty  \frac {J_{2m}( 2 g k)}{ \cosh \frac
{k}{2}} \frac {g^{2m-1}}{2(-1)^{m-1}}\left [ \frac {2^{-\frac
{5}{2}}}{g^{2m-\frac {1}{2}}}I_f- (2m-1)\frac {2^{-2}}{g^{2m}}
\right ] [1+O(g^{-1})] \, . \nonumber
\end{eqnarray}
Since the leading order terms in the sums cancel out, we are left
with the equation
\begin{equation}
\hat \rho (k)= \frac {J_0( 2 g k)}{2 \cosh \frac {k}{2}}- 2\sum
_{m=1}^{\infty}(-1)^{m-1}\frac {J_{2m}( 2 g k)}{ 2\cosh \frac
{k}{2}} [1+O(g^{-1})] \, .
\end{equation}
As in the previous subsection, the sum over $m$ is easily performed
and eliminates the Bessel functions out of the game:
\begin{equation}
\hat \rho (k)= \frac {\cos ( 2 g k)}{2 \cosh \frac {k}{2}} +
O(g^{-1}) \,  \Rightarrow \rho (u) = \frac {1}{4} \left [ \frac
{1}{\cosh \pi (u+ 2 g)}+ \frac {1}{\cosh \pi (u- 2 g)} \right ]+
O(g^{-1}) \, .
\end{equation}
This identifies the function $f(u)$ with
\begin{equation}
f(u)=\frac {1}{\cosh \pi u} \, ,
\end{equation}
furnishing again the leading density at generic $\kappa$ of previous
subsection. In particular, the value of the integral (\ref{I}) is
still given by (\ref{Ivalue}).

As further check, one can verify that the strong coupling limit
(\ref{Qrg}) of the charges actually satisfies any equation
(\ref{eqcharg}), i.e.
\begin{eqnarray}
&&Q_r= \frac { i^{r-2}}{g^{r-1}} \Bigl \{ \frac {1}{\pi
(r-1)^2 \cos \frac {\pi (r-2)}{2}}- \nonumber \\
&-& \frac {8}{\pi} g \sum _{m=1}^{\infty}  \frac {(-1)^{m-\frac
{r}{2}}}{(2m-1+r)(2m+1-r)} \frac {g^{2m+1}}{2(-1)^m} \left [ \frac
{2^{-\frac {5}{2}}}{g^{2m+\frac {3}{2}}}I-(2m+1) \frac
{2^{-2}}{g^{2m+2}}
\right ] - \nonumber \\
&-&  \frac {8}{\pi} g \sum _{m=1}^{\infty}   \frac {(-1)^{m-\frac
{r}{2}}}{(2m-1+r)(2m+1-r)} \frac {g^{2m-1}}{2(-1)^{m-1}} \left [
\frac {2^{-\frac {5}{2}}}{g^{2m-\frac {1}{2}}}I-(2m-1) \frac
{2^{-2}}{g^{2m}} \right ] \Bigr \} \, , \nonumber
\end{eqnarray}
which is easily verified upon considering the leading order of the
l.h.s. $Q_r=O(\frac{1}{g^{r-\frac{1}{2}}})$. In conclusion, the
expansion (\ref{Qrg}),
\begin{eqnarray}
Q_r&=&\left [ \frac {2^{-\frac {5}{2}}}{g^{r-\frac {1}{2}}}I-
(r-1)\frac {2^{-2}}{g^r} + O\left (\frac {1}{g^{r+\frac {1}{2}}}
\right )\right ] =
\nonumber \\
&=& \left [ \frac {2^{-\frac {3}{2}}}{g^{r-\frac {1}{2}}}\, \frac
{(2^{\frac {3}{2}}-1)}{\pi} \, \zeta \left ( \frac {3}{2} \right )-
(r-1)\frac {2^{-2}}{g^r} + O\left (\frac {1}{g^{r+\frac {1}{2}}}
\right )\right ] \, ,
\end{eqnarray}
is the exact one when $g\rightarrow +\infty$ and $\kappa=2$.
Moreover, it also yields the next-to-leading term in the
approximation defined in the previous subsection when $\kappa=1$.

In particular we can estimate the anomalous dimension within the two
first orders
\begin{equation}
\Delta= L(1+2g^2 Q_2)=L\left [2^{-\frac {1}{2}} \, \frac {(2^{\frac
{3}{2}}-1)}{\pi} \, \zeta \left ( \frac {3}{2} \right )\, g^{\frac
{1}{2}} + \frac {1}{2} + O(g^{-\frac {1}{2}})\right ] \, . \label
{Deltares}
\end{equation}

\section{Finite size equations}
\setcounter{equation}{0}

In Sections 3 and 4 we have studied the Bethe equations
(\ref{BAloops}) and the conserved charges in the anti-ferromagnetic
configuration and in the thermodynamic regime. In this Section, we
want to analyze the finite $L$ case, for slightly more general
states, and we also allow for the presence of {\it holes} in the
sequence of real Bethe roots. In this respect, it is convenient to
rewrite equations (\ref{BAloops}) in an integral form,-- called
non-linear integral equation to be distinguished by the linear one
of the thermodynamic limit \cite{FMQR}. This non-linear equation has
always revealed to be effective for the study of large but finite
size corrections, different limit regimes (strong coupling, large
size etc.) and for many other issues (cf. for instance \cite{FFGR2}
and references therein).

As far as ${\cal N}=4$ SYM is concerned, the equations
(\ref{BAloops}) at finite $L$ are reliable only up to the order
$g^{2L-2}$, because they do not take into account the wrapping
effects. Therefore, all the finite $L$ formul{\ae} presented in this
section have to be understood as relevant only up to the order
$g^{2L-2}$.

As usual \cite{FMQR}, we start from equations (\ref{BAloops}) in the
logarithmic form (\ref{logBAloops})
\begin{equation}
iL\ln \frac{X\left (u_k+\frac{i}{2}\right )} {X\left
(u_k-\frac{i}{2}\right )}= i\mathop{\sum ^M_{j=1}}_{j\neq k} \ln
\frac{u_k-u_j+i}{u_k-u_j-i} - 2 \mathop {\sum ^M_{j=1}}_{j\neq k}
\theta (u_k,u_j) +2\pi K_k\, . \label {logBAloops1}
\end{equation}
Consequently, we define two functions giving also their different
analyticity domain
\begin{equation}
\phi (x)= i \ln \frac {i+x}{i-x}=2 \arctan x \, , \, \, {\mbox
{Im}}x <1 \, ; \quad \Phi (x)=i \ln \frac {X\left (\frac
{i}{2}+x\right )}{X\left (\frac {i}{2}-x\right )} \, , \, \, {\mbox
{Im}}x <1/2 \, . \label{fidef}
\end{equation}
Upon using the relations
\begin{eqnarray}
&&i\ln \frac {x+i}{x-i}-i\ln \frac {i+x}{i-x}=-\pi \, , \nonumber \\
&&i\ln \frac {X\left (x+\frac {i}{2}\right )}{X\left (x-\frac
{i}{2}\right )}- i\ln \frac {X\left (\frac {i}{2}+x\right )}{X\left
(\frac {i}{2}-x\right )}=-\pi \, , \nonumber
\end{eqnarray}
we may recast (\ref{logBAloops1}) as
\begin{equation}
\label{logBAloops2} L\Phi (u_k) = \sum ^M_{j=1} \phi (u_k-u_j) - 2
\sum ^M_{j=1} \theta (u_k,u_j) +\pi (L-M+1+2 K_k) \, .
\end{equation}
Let us define the counting function (analytic in a strip centered on
the axis)
\begin{equation}
Z(u)=L\Phi (u) - \sum ^M_{j=1} \phi (u-u_j)+ 2 \sum ^M_{j=1} \theta
(u,u_j) \, , \label {Zdef}
\end{equation}
which renders the Bethe equations in the simple form
\begin{equation}
e^{iZ(u_k)}=-1 \, . \label {rootcondi}
\end{equation}
Now, we concentrate on the state characterised by $M$ real Bethe
roots, $u_j$, and $H$ holes, $x_h$, which are defined to satisfy
(\ref{rootcondi}), without being solutions of the original Bethe
equations (\ref{logBAloops1}). In this case, one may express a sum
on the Bethe roots for a function $f(u)$ as {\it logarithmic
indicator integral} (for a detailed discussion see \cite
{FMQR,FFGR2})
\begin{equation}
\sum ^M_{k=1} f(u_k)=- \int _{-\infty}^{\infty} \frac {du}{2\pi} \
\frac {d}{du}f(u) \ [Z(u)-2L(u)]-\sum _{h=1}^H f(x_h) \, , \label
{fsum}
\end{equation}
where we have used the short notation
\begin{equation}
L(u)= {\mbox {Im}} \ln \left [1+ (-1)^{\delta} e^{iZ(u+i0)} \right ]
\, ,
\end{equation}
and defined $\delta = L-M$ mod $2$. In particular, we will be
interested in the eigenvalues of the conserved charges
\begin{equation}
{\cal Q}_r=\sum ^M_{k=1} q_r(u_k)=- \int _{-\infty}^{\infty} \frac
{du}{2\pi} \ \frac {d}{du}q_r(u) \ [Z(u)-2L(u)] -\sum _{h=1}^H q_r
(x_h) \, , \label {qrsum}
\end{equation}
which scale to the thermodynamic values $Q_r$ (\ref{Qr}) according
to
\begin{equation}
\lim _{L\rightarrow \infty }\frac{{\cal Q}_r}{L}= Q_r \, .
\label{qrlimit}
\end{equation}
Remarkably (\ref {fsum}) may be applied to (\ref {Zdef}) itself
\begin{eqnarray}
Z(u)&=&L\Phi (u) - \int _{-\infty}^{\infty} \frac {dv}{2\pi} \frac
{2}{(u-v)^2+1}[Z(v)-2L(v)] + \sum _{h=1}^H \phi (u-x_h)-  \nonumber \\
&-& 2 \int _{-\infty}^{\infty} \frac {dv}{2\pi} \ \frac {d}{dv}
\theta (u,v) \ [Z(v)-2L(v)] -2\sum _{h=1}^H \theta (u,x_h)\, .
\nonumber
\end{eqnarray}
At this stage, it is customary to pass on to the Fourier space and
it is possible to go through this step even here. Although it is
made rather cumbersome by the involved form of the dressing phase
\begin{equation}
\theta (u,v)=\kappa \sum _{r=2}^{\infty}\sum _{\nu =0}^{\infty}
c_{r,r+1+2\nu}(g)g^{2r+2\nu-1}[q_r(u)q_{r+1+2\nu}(v)-q_{r+1+2\nu}(u)q_r(v)]
\, ,
\end{equation}
in Fourier space the following equation shall hold
\begin{eqnarray}
&&(1+e^{-|k|})\hat Z(k)= L \frac {2\pi}{i} P\left (\frac {1}{k}\right ) e^{-|k|/2} J_0(2gk) +2 e^{-|k|} \hat L(k) + \nonumber \\
&+& \sum _{h=1}^H e^{-ikx_h}\frac {2\pi}{i} P\left (\frac
{1}{k}\right ) e^{-|k|}- 2 \kappa \int _{-\infty}^{\infty} \frac
{dv}{2\pi} \sum _{r=2}^{\infty}\sum _{\nu=0}^{\infty}
c_{r,r+1+2\nu}(g)g^{2r+2\nu-1} \cdot \nonumber \\
&\cdot & [\hat q_r(k)\frac {d}{dv} q_{r+1+2\nu}(v)-\hat
q_{r+1+2\nu}(k)\frac {d}{dv}
q_r(v)][Z(v)-2L(v)] -   \\
&-& 2{\kappa} \sum _{h=1}^H \sum _{r=2}^{\infty}\sum
_{\nu=0}^{\infty} c_{r,r+1+2\nu}(g)g^{2r+2\nu-1}[\hat q_r(k)
q_{r+1+2\nu}(x_h)-\hat q_{r+1+2\nu}(k) q_r(x_h)]  \nonumber \, .
\end{eqnarray}
Inspired by the thermodynamic case, we simplify a little this
relation by introducing the charges ${\cal Q}_r$
\begin{eqnarray}
\hat Z(k)&=& L \frac {2\pi}{i} P\left (\frac {1}{k}\right ) \frac
{J_0(2gk)}{2\cosh \frac {k}{2} }  + \frac {2}{1+e^{|k|}} \hat
L(k) + \sum _{h=1}^H e^{-ikx_h} \frac {2\pi}{i}  P\left (\frac {1}{k}\right ) \frac {1}{1+e^{|k|}} +\nonumber  \\
&+&\frac {\kappa}{\cosh \frac {k}{2}} \sum _{r=2}^{\infty}\sum
_{\nu=0}^{\infty} c_{r,r+1+2\nu}(g)g^{2r+2\nu-1} \Bigl [\frac {2\pi
}{g^{r-1}}\frac {1}{i^{r-2}}\frac
{J_{r-1}(2gk)}{k}{\cal Q}_{r+1+2\nu}- \nonumber \\
&-& \frac {2\pi }{g^{r+2\nu}}\frac {1}{i^{r+2\nu-1}}\frac
{J_{r+2\nu}(2gk)}{k}{\cal Q}_r\Bigr ] \, , \label {Zeq2}
\end{eqnarray}
and also the explicit form (\ref {qrk}) of the Fourier transform of
the charge densities, $\hat q_r(k)$. A very crucial difference of
this non-linear integral equation from the others in the literature
may be stated in the presence of $Z(u)$ in infinite many place, i.e.
all the charges ${\cal Q}_r$ (\ref{qrsum}).

Concerning the latter, we may correct the thermodynamic system of
equations in case of finite $L$. In fact, we first rewrite the
expressions (\ref{fsum}) in terms of Fourier transforms
\begin{eqnarray}
{\cal Q}_s&=&-\int _{-\infty}^{\infty} \frac {dk}{4\pi^2} \hat
{q^{\prime}_s} (-k) [\hat Z(k) -2 \hat L(k) ] -\sum _{h=1}^H q_s(x_h)= \nonumber \\
&& = \int _{-\infty}^{\infty} \frac {dk}{2\pi} \frac
{i^{3+s}}{g^{s-1}}\frac {J_{s-1}(2gk)}{e^{ \frac {|k|}{2}}} [ \hat
Z(k) - 2 \hat L(k) ] -\sum _{h=1}^H q_s(x_h) \, .
\end{eqnarray}
Then we insert relation (\ref {Zeq2}) into this expression to obtain
the corrections
\begin{eqnarray}
&&{\cal Q}_s=\frac {i^{s+2}}{g^{s-1}} \Bigl [ L \, \int
_{-\infty}^{\infty} dk \frac {J_{s-1}
( 2 g k) J_0( 2 g k)}{k (e^{|k|}+1)} + \nonumber \\
&+& 2\kappa  \sum _{r=2}^{\infty} \sum _{\nu =0}^{\infty}
c_{r,r+1+2\nu}(g) (-1)^{1+\nu} \int _{-\infty}^{\infty} dk \frac
{J_{s-1} ( 2 g k) J_{r-1}( 2 g k)}{k (e^{|k|}+1)}  \frac {
g^{r+2\nu}}{ i^{r+2\nu-1}} {\cal
Q}_{r+2\nu+1}  \nonumber \\
&+&  2\kappa  \sum _{r=2}^{\infty} \sum _{\nu =0}^{\infty}
c_{r,r+1+2\nu}(g) (-1)^{1+\nu} \int _{-\infty}^{\infty} dk \frac
{J_{s-1} ( 2 g k) J_{r+2\nu}( 2 g k)}{k (e^{|k|}+1)} \frac {
g^{r-1}}{ i^{r-2}} {\cal
Q}_{r} \Bigr ] - \nonumber \\
&-& \int _{-\infty}^{\infty} \frac {dk}{2\pi} \frac
{i^{3+s}}{g^{s-1}}\frac {J_{s-1}(2gk)}{\cosh \frac {k}{2}} \hat L(k)
-\sum _{h=1}^H \int _{-\infty}^{\infty} \frac {dk}{2k} e^{-ikx_h}
\frac {i^{2+s}}{g^{s-1}}\frac {J_{s-1}(2gk)}{\cosh \frac {k}{2}} \,
. \label{eqcharg2}
\end{eqnarray}
This relation is exact and, at least in principle, may be efficient
in the analysis of the conserved charges, though now $Z(u)$ appears.

\section{Outlook}

In this paper we have pointed out some aspects of the (long-range)
asymptotic Bethe equations with dressing factor for the $SU(2)$
sector of planar ${\cal N}=4$ SYM. In the specific case of the
anti-ferromagnetic state in the thermodynamic limit, we have written
an integral equation for the density of Bethe roots and solved it in
the leading strong coupling limit. Upon integration, this density
gives an approximation for (the eigenvalues of) the commuting
charges on this state. Nevertheless, for these charges we have also
derived an exact linear system (of algebraic equations): we do think
that its investigation may clarify many points, especially about the
strong coupling expansion. As a mathematical curiosity, we have
observed that these equations are exactly solved in the strong
coupling limit if one doubles the dressing factor of \cite{BHL,BES}.
At present, this mathematical simplification has no physical
meaning, but the exactness of the solution at this special value of
the dressing. As well known in (integrable) statistical field
theory, the linear integral equation converts into a non-linear one
when the size $L$ becomes finite: in this respect we have shown that
this case does not make any exception, though the specific equation
is rather involved. Moreover, we have focused our attention at most
to the hole 'excitations' on the anti-ferromagnetic configuration
and we have derived a finite-size corrected system of equations for
the charges. Eventually, we should be able to give simple
generalisations of the aforementioned results to the other
states/operators along the lines of \cite{FMQR}, including {\it
complex pairs}.

\medskip

{\bf Acknowledgements} We have benefited from discussions with
N.Beisert, D. Bombardelli, F. Ravanini, M. Staudacher and K.Zarembo.
We acknowledge the INFN grant "Iniziativa specifica TO12/PI14" for
travel financial support. M.R. thanks INFN-Bologna and the
Department of Physics in Bologna for generous hospitality.

\end{document}